\documentclass[twocolumn,pra,showpacs]{revtex4}
\usepackage{graphicx}
\usepackage[dvips]{color}

\begin{document}

\newcommand{\be}{\begin{equation}}
\newcommand{\ee}{\end{equation}}
\newcommand{\bea}{\begin{eqnarray}}
\newcommand{\eea}{\end{eqnarray}}


\title{Making Cold Molecules by Time-dependent Feshbach Resonances }
\author{ Paul S. Julienne}
\author{Eite Tiesinga}
\affiliation{Atomic Physics Division, National Institute of Standards
and Technology, Gaithersburg, MD 20899-8423}
\author{Thorsten K\"{o}hler}
\affiliation{Clarendon Laboratory, Department of Physics,  
University of Oxford, Parks Road, Oxford, OX1 3PU, United Kingdom}

\begin{abstract}
Pairs of trapped atoms can be associated to make a diatomic molecule using a time dependent magnetic field to ramp the energy of a scattering resonance state from above to below the scattering threshold.   A relatively simple model, parameterized  in terms of the background scattering length and resonance width and magnetic moment, can be used to predict conversion probabilities from atoms to molecules.  The model and its Landau-Zener interpretation are described and illustrated by specific calculations for $^{23}$Na, $^{87}$Rb, and $^{133}$Cs resonances.  The model can be readily adapted to Bose-Einstein condensates.  Comparison with full many-body calculations for the condensate case show that the model is very useful for making simple estimates of molecule conversion efficiencies.
\end{abstract}

\date{\today}
\maketitle

\section{Introduction}\label{intro}

The production and characterization of cold, trapped molecules and quantum degenerate molecular gases is an important contemporary challenge in physics.  Considerable progress towards this goal has occurred recently, using the properties of magnetically tunable Feshbach resonance states to convert atoms in an ultracold Bose~\cite{Donley02,Herbig03,Chin03,Durr03,Xu03} or Fermi~\cite{Regal03,Srecker03,Cubizolles03,Jochim03,Zwierlein03} gas to bound state molecules comprised of a pair of atoms.  This paper describes a simple model for understanding the basic molecular physics underlying the conversion of atom pairs to diatomic molecules by a time-dependent ramp of a scattering resonance across the collision threshold.  We have in mind the case of two atoms confined in an individual cell of an optical lattice, as well as a trapped quantum degenerate gas of bosons.  Although we do not treat quantum degenerate Fermi gases here, the 2-body physics we discuss is applicable to either fermionic or bosonic atoms.  The molecular physics of threshold scattering of alkali species is now well-characterized~\cite{Burnett02}.  We will show how the model proposed by Mies {\it et al.}~\cite{Mies00} gives a very simple way to estimate molecule production rates in terms of only three parameters that characterize the molecular physics and scattering.  We also show that the model makes predictions that are quantitatively similar to the results of full many-body calculations for a trapped Bose-Einstein condensate (BEC) using the methods of K\"ohler {\it et al.}~\cite{Koehler02,Koehler03,Koehler03b,Goral03}.

Figure 1 illustrates a Feshbach ramp by showing the time-dependent energy levels for a case to be discussed below using the model of Ref. ~\cite{Mies00}.  The idea is to take a quasibound molecular resonance state whose energy $\epsilon_n$ is well above the threshold energy $\epsilon=0$ for the two-body collision, and sweep the resonance state energy below threshold by a time-dependent change in its position.  Here we consider the case where this is done by tuning a magnetic field, although it is also possible to use an optical frequency electromagnetic field.  If the sweep rate is sufficiently slow, then the initially populated ground state $|v=0\rangle$ of a trapped atom pair adiabatically evolves to an atom pair in the molecular resonance state $|n\rangle$, now located below threshold and stable with respect to decay into two atoms.  For two atoms in a trap, the transfer of population to the final states is accurately described by a Landau-Zener model.  In general, the molecular physics of scattering resonances for alkali species is now well-understood, with remarkable agreement between coupled channels theoretical models and measurement~\cite{Leo00,Vuletic00,Marte02}.  The molecular physics associated with an isolated resonance is characterized by three key parameters: the background scattering length $A_{bg}$ away from resonance, the resonance width $\Delta_n$, and the magnetic moment difference between the resonant state $|n\rangle$ and the separated atoms proportional to $s_n=\partial{\epsilon_n}/\partial{B}$. These parameters can, in principle, be determined from either calculation or measurement.  In the Landau-Zener model the molecule formation probability depends only on the product $A_{bg}\Delta_n$, and not on $s_n$.

\begin{figure}[!htb]
\centering
\includegraphics[scale=0.3]{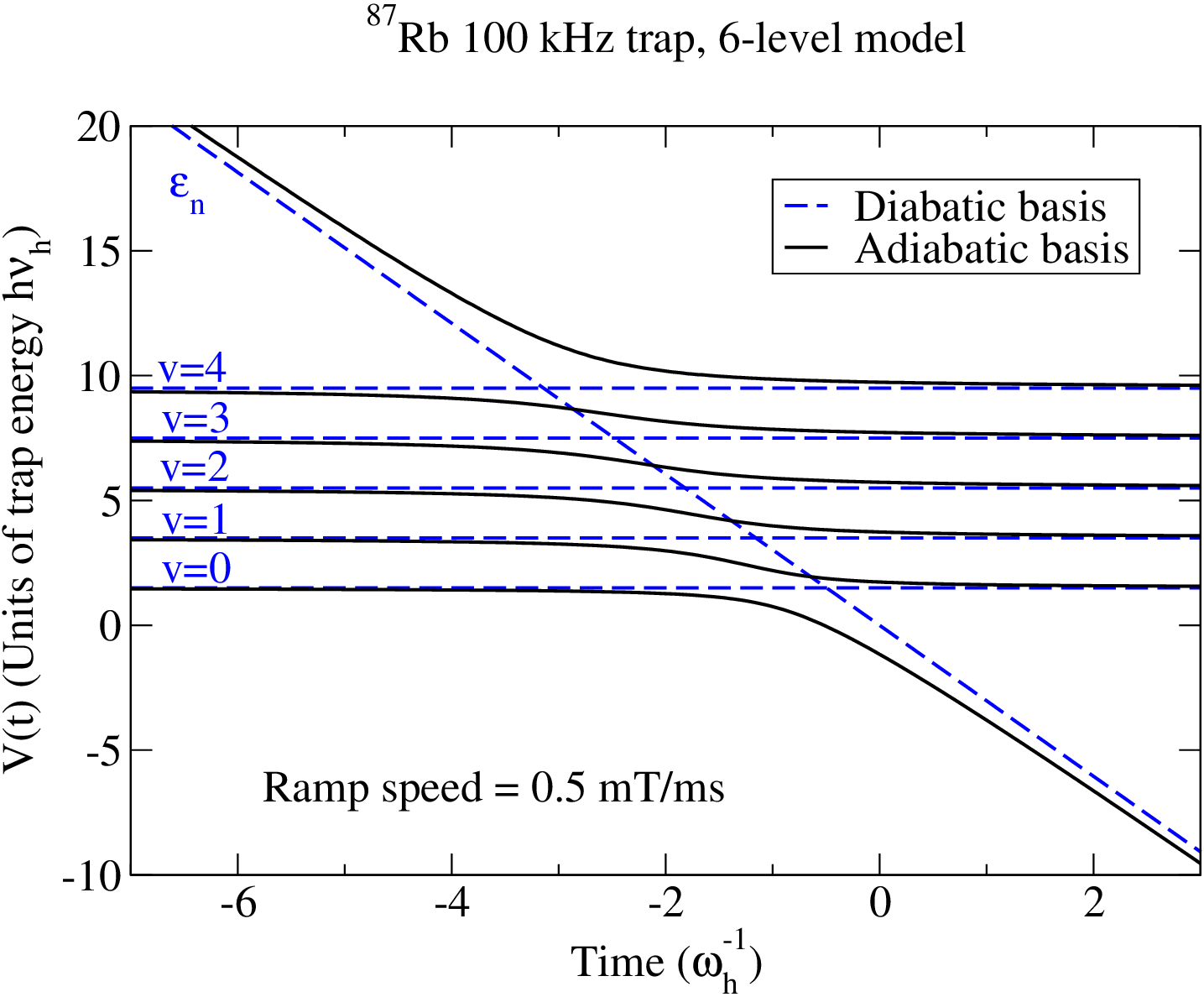}
\caption{Energy levels for a 6-level model (one tunable resonant state and five trap levels, omitting higher ones) of a time-dependent Feshbach resonance ramp for two $^{87}$Rb F$=1$, M$=+1$ atoms in a $\omega_h=2\pi(100$ kHz$)$ isotropic harmonic trap.  Energies are relative to the energy of two separated atoms, which is taken to define $\epsilon=0$.  The dashed lines show the diabatic (or "bare") energy levels before the interactions between the trap levels $|v\rangle$ and bound state $|n\rangle$ are turned on.   The solid lines show the adiabatic (or "dressed") eigenenergies of the full Hamiltonian.  The diabatic resonance state energy $\epsilon_n$ is ramped linearly in time by changing the magnetic field $B$ at a speed of 0.5 T/s, such that $\epsilon_n=0$ at time $t=0$.  The time scale is in units of $\omega_h^{-1}=1.16\mu$s.  A pair of atoms initially in the ground trap state $|v=0\rangle$ for time $t\ll -\omega_h^{-1}$ can evolve adiabatically to the bound molecular state $|n\rangle$ for $t\gg \omega_h^{-1}$.  The model  resonance state $|n\rangle$ has a width $\Delta_n=0.017$ mT and a slope (see text) $s_n/h=38$ MHz/mT.  This case corresponds to an actual $^{87}$Rb resonance near 100.76 mT~\cite{Marte02}. }
\label{Fig1}
\end{figure}

Ref.~\cite{Mies00}  showed the connection between the conventional two-body scattering picture of the resonance interacting with a continuum and the corresponding properties involving discrete quantized levels in a trap.  The paper also showed how to modify the model when the atoms are in a Bose-Einstein condensate (BEC).  Here we will summarize the key results of Ref.~\cite{Mies00}, following its notation, and extend and apply the method to several specific cases.  Section ~\ref{section1} considers the collision of two atoms in free space, influenced by a single tunable resonance level.  Section ~\ref{section2} shows how to adapt this picture to time-dependent dynamics in a trap when the magnetic field ramps the resonance energy across the collision threshold. Section ~\ref{section3} describes the many-body adaptation needed for resonance ramps in a Bose-Einstein condensate.  Section ~\ref{section4} gives results for specific cases, and compares the simple model predictions with those of full many-body calculations.

\section{Threshold resonant collisions in free space}\label{section1}

Mies {\it et al.}~\cite{Mies00} showed how the full coupled channels quantum scattering problem in the case of a set of closed channels with isolated resonance levels interacting with a single scattering continuum can be reduced to an effective two-channel configuration interaction (CI) problem.  The wavefunction at collision kinetic energy $\epsilon$ in a magnetic field of strength $B$ is represented by the following CI wavefunction:
\begin{equation}
\Psi_{n}(\epsilon,B,R) \equiv |1\rangle F_{1}(\epsilon,R,B) + 
                       |n\rangle \phi_n(R) A_{n}(\epsilon,B) \,.
\label{CIPsi}
\end{equation}
Here $R$ is interatomic separation, $|1\rangle$ represents the internal spin variables of the open collision channel, $F_1(\epsilon,R,B)$ is the open-channel scattering function, $|n\rangle \phi_{n}(R)$ represents the resonant closed-channel bound state with internal spin state $|n\rangle$ and motional state $\phi_n(R)$, and  $A_{n}(\epsilon,B)$ is the amplitude of the resonant state.  The wavefunction (\ref{CIPsi}) is the solution to the Hamiltonian
\begin{eqnarray}
H_{n}(B)&=&|1\rangle\langle1|\{T+U_{bg}(R)\}+ \nonumber \\
    && |n\rangle\langle n|\{T+U_{n}(B,R)\}+ \nonumber \\ 
    && \{|1\rangle\langle n|+|n\rangle\langle1|\}W_{n,1}(R)
\label{CIHam}
\end{eqnarray}
where $T$ is the radial kinetic energy operator.  The $|1\rangle$ and $|n\rangle$ basis states are "bare" or diabatic states before the interchannel coupling $W_{n,1}(R)$ is taken into account.  Ref.~\cite{Mies00} demonstrated how the potentials $U_{bg}(R)$ and $U_n(R)$ and the coupling operator $W_{n,1}(R)$ can be replaced by convenient  effective potentials, or pseudopotentials, and an effective coupling operator.  These effective potentials and operator reproduce all near-threshold scattering properties. 

The amplitude $A_{n}(\epsilon,B)$ of the resonant state is independent of $R$, and the bound state radial vibrational function $\phi_n(R)$ is a solution to the single channel diabatic bound state eigenvalue problem:
\begin{equation}
[-\frac{\hbar^2}{2\mu}\frac{d^2}{dR^2}+U_{n}(B,R)]\phi_{n}(R)=
     \epsilon_{n}(B) \phi_{n}(R)  \,,
\label{Eqphi}
\end{equation} 
where $U_n(B,R)$ is the closed channel effective potential and $\mu$ is the reduced mass of the pair of atoms.  The function $ \phi_{n}(R)$ is just an ordinary vibrational wavefunction, the shape of which is independent of $B$.  

The effect of the magnetic field is merely to shift the position of the bound state energy $\epsilon_{n}(B)$.   Let $B_{n}$ define the resonance field where the diabatic bound state crosses the open channel threshold at $\epsilon=0$, i. e., $ \epsilon_{n}(B_{n})=0$.  Note that we set the open channel threshold energy, that is, the energy of the two separated atoms,  to be 0 independent of $B$.  This emphasizes that what is important is the energy of the resonance state relative to the energy of the two separated atoms.  Then for $B$ close to $B_{n}$, 
\begin{equation}
\epsilon_{n}(B)= s_n(B-B_{n})
\label{Eres}
\end{equation}
where the slope $s_n={\partial \epsilon_{n}(B)}/{\partial B}$ is a constant equal to the difference in magnetic moments of the resonance state and two separated atoms.  Assuming that the magnetic moments do not vary over the small range of $B$ considered, varying the magnetic field simply causes the resonance energy to ramp linearly across threshold while $ |n\rangle \phi_{n}(R)$ remains unchanged.  

The open channel scattering solution has the asymptotic form at large $R$:
\begin{equation} 
F_{1}(\epsilon,R,B)\rightarrow \sqrt{\frac{2\mu}{\pi\hbar^{2}k}} \sin(kR+\xi(B))
\label{F1asym} 
\end{equation} 
where $k=\sqrt{2\mu \epsilon/\hbar^2}$ and
\be
\xi(B)=\xi_{bg} + \xi_n(B) \,,
\label{Xires}
\ee
where $\xi_{bg}$ and $\xi_n(B)$ are the respective background and resonance contributions to the phase shift $\xi$.
When $B$ is such that the resonance energy is tuned far from threshold, the effect of the resonance is negligible, and the scattering reduces to the background scattering by the open channel effective potential $U_{bg}(R)$:
\begin{equation}
[-\frac{\hbar^2}{2\mu}\frac{d^2}{dR^2}+U_{\rm bg}(R)]\phi_{\epsilon}(R)=
     \epsilon \phi_{\epsilon}(R)  
\label{OpenSE}
\end{equation} 
where $\epsilon>0$ is the collision kinetic energy.  The asymptotic solution of Eq. (\ref{OpenSE}) has the form of Eq. (\ref{F1asym}) with $\xi = \xi_{bg}$.  A key collision parameter is the background scattering length $A_{bg}$:
\be
  A_{bg} =  \lim_{k\to0}\frac{-\xi_{bg}}{2k}   \,.
\label{BgSlen}
\ee

There is also a resonance contribution to the phase shift, which is important when $B$ is tuned so that $\epsilon_{n}(B)$ is near threshold:
\begin {equation} 
\tan{\xi_{n}(\epsilon,B)}=\frac{\Gamma_{n}(\epsilon)} 
{2[\epsilon-\epsilon_{n}(B)-\delta\epsilon_n(B)]}\,,
\label{ResPhase}
\end {equation}
where $\Gamma_{n}(\epsilon)$ is the decay width and $\delta\epsilon_n(B)$ a shift from the "bare" resonance crossing point (see below).  The decay width is given by the Fermi golden rule:
\begin{equation}
\Gamma_{n}(\epsilon) \equiv 	2\pi | V_n(\epsilon)|^2 =
     2\pi|\langle \phi_{n}|W_{n,1}|\phi_{\epsilon}\rangle|^{2} \,.
     \label{CIwidth}
\end{equation}
The Wigner threshold law for $s$-waves requires that $\Gamma_{n}(\epsilon) \propto k$ as $k \to 0$.

When $k$ is very small, such that $kA\ll1$, the definition of the scattering length $A$,
\be
 A(B) =  \lim_{k\to0}\frac{-\xi(B)}{2k} = A_{bg} + A_n(B) \,.
\label{Slen}
\ee
using the $k\to0$ limit of Eq.~(\ref{ResPhase}), gives the common formula
\be
 A(B)=A_{bg}\left ( 1 - \frac{\Delta_n}{B-B_n-\delta B_n} \right ) \,,
\label{SlenB}
\ee
where
\be
\Delta_n s_n=  \frac{\Gamma_n(\epsilon)}{2 k A_{bg}} \,.
\label{Delta}
\ee
The width $\Delta_n$ is a constant independent of $k$ close to threshold, since the ratio $\Gamma_n(\epsilon)/k$ is independent of $k$ due to the threshold law.  Note that $A_{bg}$ and $s_n$ can be positive or negative, so that $\Delta_n$ has the same sign as the product $A_{bg}s_n$.

Using the methods of Ref.~\cite{Raoult00}, the shift can be shown to be 
\be
  \delta B_n= \frac{r(1-r)}{1+(1-r)^2} \Delta_n \,,
  \label{Shift}
\ee
where $r=A_{bg}/\bar{A}$, and $\bar{A}=0.477989 (2\mu C_6/\hbar^2)^{1/4}$ is the mean scattering length defined by Gribakin and Flambaum~\cite{Gribakin93}, for a collision of two atoms with reduced mass $\mu$ in a van der Waals potential with coefficient $C_6$.  The trap CI model discussed in the next Section implicitly includes the shift, if a complete set of states for the background potential $U_{bg}$ is included in the basis.  It is especially important if $A_{bg}$ is large and positive to include the last bound state of $U_{bg}$ in the basis, since it can make a large contribution to the shift (see Section~\ref{section4.3} below).  However, the final populations in the CI model after the ramp are completely independent of the shift, although the detailed dynamics during the ramp can be strongly modified when the shift is taken into account.  For the purpose of this paper, which is primarily concerned with the final populations, we can safely ignore the shift.  

\section{Threshold resonant collisions in a trap}\label{section2}

The free-space picture described in Section \ref{section1} needs to be modified when a collision occurs in a trap, where the atoms are no longer free to separate to arbitrarily large $R$.  Their motion is confined to the trapping volume, and the free scattering states must be replaced by the quantized eigenstates of the interacting atoms in the trap\cite{Tiesinga00,Bolda02,Blume02}.  The center of mass and relative motion of the two atoms are uncoupled for harmonic traps; the weak coupling between them for actual anharmonic traps that are used for atom trapping is generally small and can be neglected\cite{Bolda03}.  The potential $U_{bg}(R)$ in Eq. (\ref{OpenSE}), which describes the relative motion of the atoms, is replaced by 
\be
U_{bg}(R)+U_{\mathrm trap}(R) \,, 
\label{TrapU}
\ee
which includes the trapping potential.   The continuous spectrum of collision energies is replaced by a discrete spectrum with trap energies $\epsilon_v$ and trap eigenfunctions $\phi_v(R)$.  Here we take $v=0,1,...$ to represent the eigenvalues $\epsilon_v>0$ of trap-like states, and $v=-1, -2, ...$ to represent the bound dimer states with $\epsilon_v<0$.  These dimer levels of the $U_{bg}$ potential exist even when the trap is not present.  The continuum function $F_1$ in Eq. (\ref{CIPsi}) is replaced by a discrete basis expansion over trap states, with time-dependent coefficients since we will be considering time-dependent Hamiltonians.  The discrete CI wavefunction is:
\begin{equation}
\Psi_{n}(t)=|1\rangle\sum_{v}\phi_{v}(R)C_{v}(t)+|n\rangle\phi_{n}
(R)A_{n}(t) \,.
\label{TdepCI}
\end{equation}
Although this equation is written using the "bare" or diabatic basis, it is also possible to diagonalize the Hamiltonian matrix and produce a "dressed" or adiabatic basis in which the wave function can be expanded.  We will set up the coupled equations in the diabatic basis, since that is very simple.  Below we will show populations calculated in both the diabatic and the adiabatic basis sets.

We assume we start at time $t_0$ with all the population in the ground state of the trap, $v=0$, with the magnetic field $B_0$ chosen so that $\epsilon_n(B_0)>0$ is above threshold.  Then the magnetic field is changed linearly in time so that the resonance level is ramped below threshold:
\begin{equation}
   B(t)=\left\{ \begin{array}{cl} 
                  B_{0}                                 & t<t_0 \\
	          B_{0}+\frac{\partial B}{\partial t} (t-t_0) & t\geq t_0 
	        \end{array} \right.
	 \label{Bt}
\end{equation}
The sign of $\partial{B}/\partial{t}$ depends on the sign of $s_n$; clearly $s_n \partial{B}/\partial{t}<0$ for downward ramps.  We only use examples in this paper for which $s_n>0$ and $\partial{B}/\partial{t}<0$, and we refer to $|\partial{B}/\partial{t}|$ as the speed of the ramp.  Inserting the CI wavefunction (\ref{TdepCI}) in the time-dependent Schr\"odinger equation gives the coupled set of equations that govern the evolution of the CI coefficients:
\begin{eqnarray}
i\hbar \dot{A_{n}}=&[\epsilon_{n}(B_{o})+s_n \frac{\partial B}{\partial 
t}t] A_{n}&+\sum_{v}V_{n,v}C_{v} \nonumber \\
i\hbar \dot{C_{v}}=& \epsilon _{v} C_{v}&+V_{n,v}A_{n} \,,
\label{CIeqs}
\end{eqnarray}

It is a straightforward procedure to solve Eqs.~(\ref{CIeqs}) once we know the coupling matrix elements
\begin{equation}
    V_{n,v}=\langle\phi_{n}|W_{n,1}(R)|\phi_{v}\rangle \,.
    \label{Vnv1}
\end{equation} 
These can be evaluated numerically for bound levels below threshold with $v<0$, whereas for trap levels with $v\ge0$ they can be found readily from the free-space matrix elements
\begin{equation}
\langle\phi_{n}|W_{n,1}(R)|\phi_{v}\rangle \sqrt{\frac{\partial 
v}{\partial \epsilon_{v}}}= 
\langle\phi_{n}|W_{n,1}(R)|\phi_\epsilon\rangle
\label{FB-BB}
\end{equation}
where $\partial v/\partial \epsilon_{v}$ measures the density of trapped states in the 
vicinity of $\epsilon_{v}\approx \epsilon_{n}$.  Using Eq. (\ref{CIwidth}), we find~\cite{Mies00}
\be
 V_{n,v}=\left ( \frac{\Gamma_n(\epsilon_v)}{2\pi}\frac{\partial \epsilon}{\partial v} \right )^{1/2}\,.
\label{Vnv2}
\ee

There are two cases we will consider.  For a harmonic potential with frequency $\omega_h$, and harmonic oscillator length $L_h=\sqrt{\hbar/\mu \omega_h}$, the density of states is a constant, ${\partial \epsilon}/{\partial v} =2\hbar\omega_h$. If $L_h$ is much larger than $A_{bg}$, the small energy shifts due to atom interactions~\cite{Tiesinga00} can be ignored, and the spherical trap energy levels are $\epsilon_v=\hbar\omega_h(\frac{3}{2}+2v)$.  The other case is a spherical box of length $L_b$.  The box eigenvalues are $\epsilon_v=\hbar\omega_b(v+1)^2$, with $\hbar\omega_b=(\hbar^2/2\mu)(\pi/L_b)^2$, and the density of states is ${\partial \epsilon}/{\partial v}=2\hbar\omega_b (v+1)$.

Using these properties and Eq. (\ref{Delta}), we find for the harmonic and box cases:
\be
 V^h_{nv}=\hbar\omega_h \sqrt{2\sqrt{3}/\pi} \sqrt{|a_h\delta_h|} (1+4v/3)^{1/4} \,,
\label{Vnvh}
\ee
\be
 V^b_{nv}=\hbar\omega_b \sqrt{2} \sqrt{|a_b\delta_b|} (1+v) \,,
\label{Vnvb}
\ee
where $a_i=A_{bg}/L_i$ and $\delta_i=\Delta_n s_n/(\hbar\omega_i)$ for $i=h,b$ are dimensionless coupling parameters.  If the last bound state in the potential is not too close to threshold, that is, if the scattering length is not too large in magnitude, the $V_{n,-1}$ coupling to the last bound state $v=-1$ can be ignored in Eq. (\ref{CIeqs}).  Thus, a knowledge of the resonance parameters $A_{bg}$, $\Delta_n$ and $s_n$ is sufficient to calculate the time dependent dynamics for a given trap using Eqs. (\ref{CIeqs}).

Mies {\it et al.}~\cite{Mies00} showed that the populations of the ground trap level $p_0$ and the resonant level $p_n$ after a linear ramp of resonance energy across threshold are accurately described by a Landau-Zener (LZ) curve crossing formula.  Assuming an initial population of unity in the ground trap level at $t=t_0$, the final populations are:
\begin{eqnarray}
p_0&=& e^{-A^{\mathrm LZ}_{n,0}} \\
p_n&=& 1 - e^{-A^{\mathrm LZ}_{n,0}}
\end{eqnarray}
where
\be
A^{\mathrm LZ}_{n,0} = \frac{2\pi |V_{n,0}|^{2}}{\hbar \left | s _{n}
\frac{\partial B}{\partial t}\right |} =\frac{4\pi f_i \omega_i}{L_i}\left | \frac{A_{bg}\Delta_n }{\frac{\partial B}{\partial t}} \right | \,,
\label{ALZ}
\ee
and $f_i=1$ if $i=b$ and $f_i=\sqrt{3}/\pi$ if $i=h$.  Note that the LZ parameter depends on the resonance only through the ratio  $A_{bg}\Delta_n/\frac{\partial B}{\partial t}$ and does not depend on $s_n$.  The LZ populations are only asymptotic ones, for long time $t$ after the crossing.  It is necessary to integrate the coupled Eqs.~(\ref{CIeqs}) in order to obtain the time-dependent populations.

The number of atoms lost from the initially populated $v=0$ trap ground state level does not depend on the direction of the ramp, that is, whether it starts out above or below threshold.  If the ramp starts with the resonance level above threshold and moves it below, then the loss of atoms from the $v=0$ level represents formation of stable molecules in the resonance level.  If the ramp starts with the resonance level below threshold and moves it above into the scattering continuum, then the loss of atoms from the $v=0$ trap level results in  heating, that is, population in levels with $v>0$.  

\begin{figure}[!htb]
\centering
\includegraphics[scale=0.3]{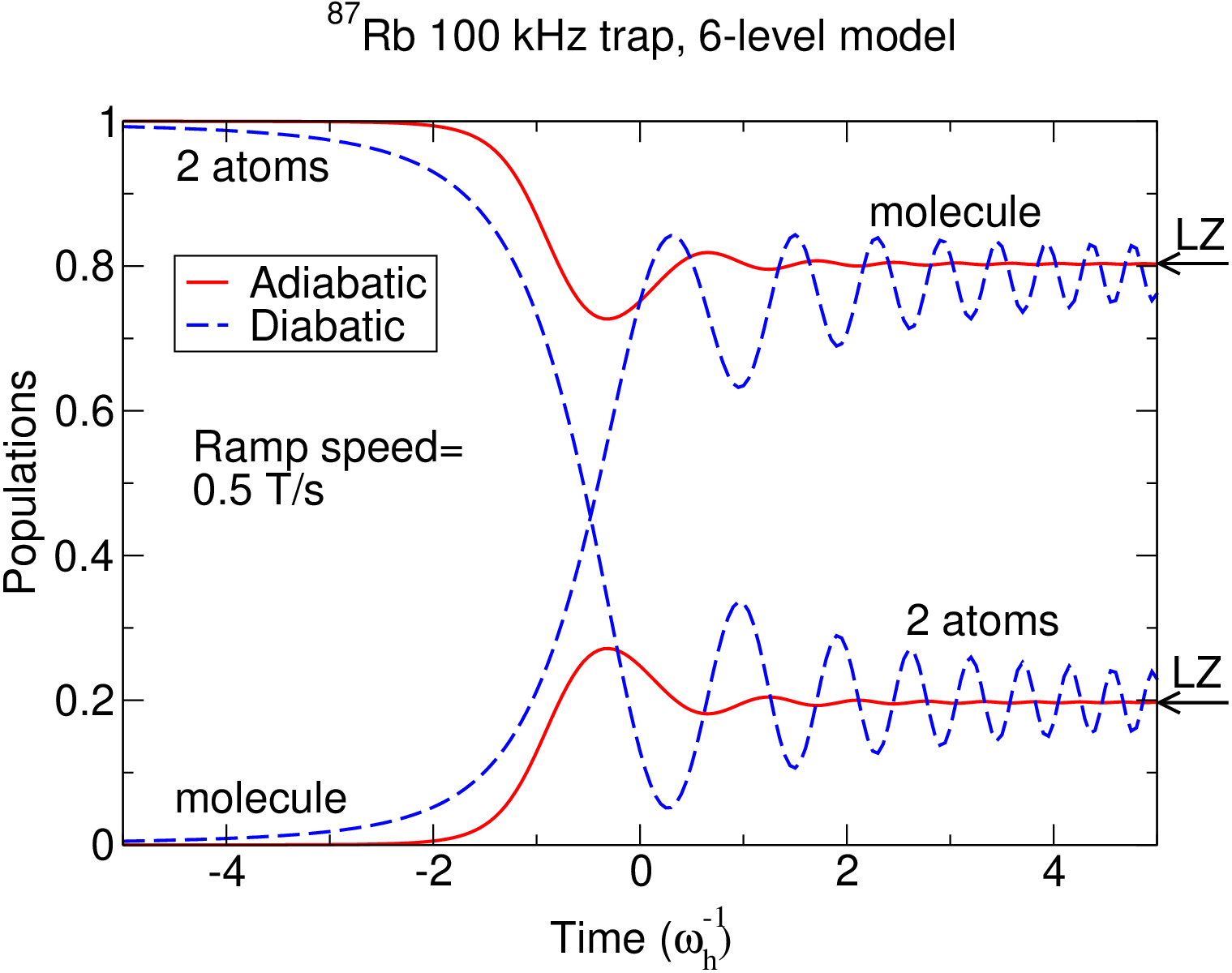}
\caption{Time dependent populations projected in the diabatic and adiabatic basis sets for the model problem shown in Figure ~\ref{Fig1} for a ramp speed of 0.5 mT/ms.  The ramp starts at $t_0=-25 \omega_h^{-1}$.  The diabatic levels shown are the "bare" ground state $v=0$ (2 atoms) and the resonance level $n$ ("bare" molecule).  The adiabatic populations shown are for the lowest two adiabatic states in Fig.~\ref{Fig1}.  Initially, the lowest adiabatic level is almost indistinguishable from the "bare" ground trap level.  This level adiabatically evolves to become nearly indistinguishable from the "bare" resonance level for times long after the threshold crossing at $t=0$. The initially unpopulated $v=1$ trap level adiabatically evolves to become the $v=0$ ground trap level long after the crossing.  Although the final state populations evolve to the Landau Zener limit, the transient populations near $t=0$ may in the general case depend on the size of the CI basis set, especially in cases where the background scattering length is large in magnitude.  The populations strongly depend on the basis in which they are projected.}
\label{Fig2}
\end{figure}

As a specific example, Fig.~\ref{Fig2} shows the time dependent populations calculated by integrating the coupled CI Eqs.~(\ref{CIeqs}) for the model shown in Fig.~\ref{Fig1}.  As the resonance nears the crossing at time $t=0$, the populations undergo transient oscillations and settle down at late time at the values given by the Landau-Zener model.  If we apply the Demkov-Osherov~\cite{Demkov67,Yurovsky99} linear multiple curve crossing model to this case, for a ramp starting at $t=-\infty$, there is exactly zero final population transfered to excited trap levels, i.e, those with $v\ge1$; such populations can only be excited transiently around the time of the crossing.  All final population either remains in $v=0$ or is transfered to the resonance state $n$.  Ref.~\cite{Mies00} showed that for either downward or upward ramps, the transition probablilites could be calculated accurately for one or for a sequence of curve crossings using the Landau-Zener expressions.  This is because for a linear set of curve crossings, the Demkov-Osherov model is an exact representation of the quantum dynamics.   

\begin{figure}[!htb]
\centering
\includegraphics[scale=0.3]{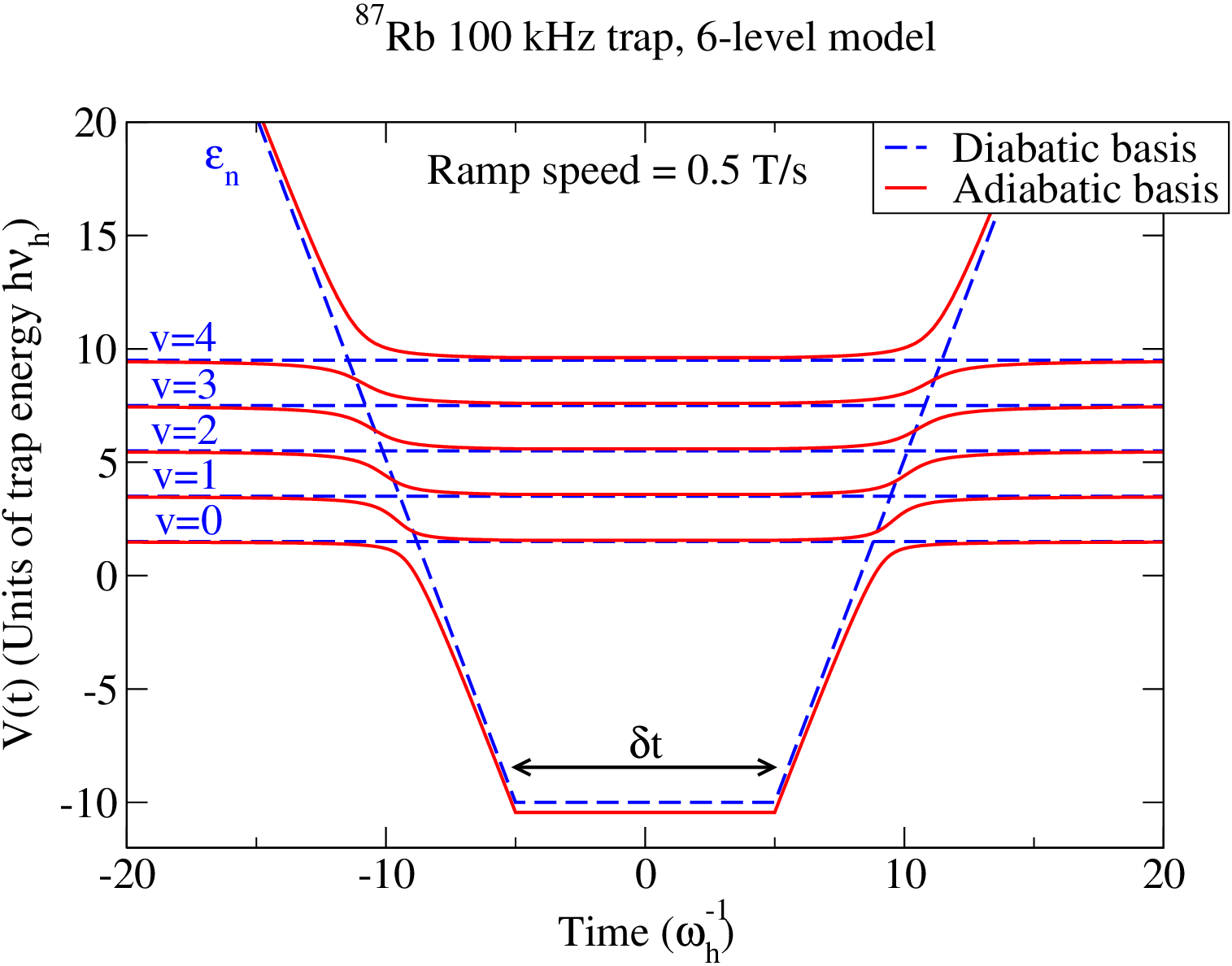}
\caption{Energy levels for the same 6-level model described in Fig.~\ref{Fig1}, except with a backwards ramp of the resonance level above threshold at the same ramp speed after a hold time $\delta t$ at a constant magnetic field, where the diabatic resonance energy is held at an energy of $-10\hbar\omega_h$.  If the ground trap state $v=0$ is initially populated, then final populations will be distributed among the trap final states with $v\ge0$. }
\label{Fig3}
\end{figure}

\begin{figure}[!htb]
\centering
\includegraphics[scale=0.3]{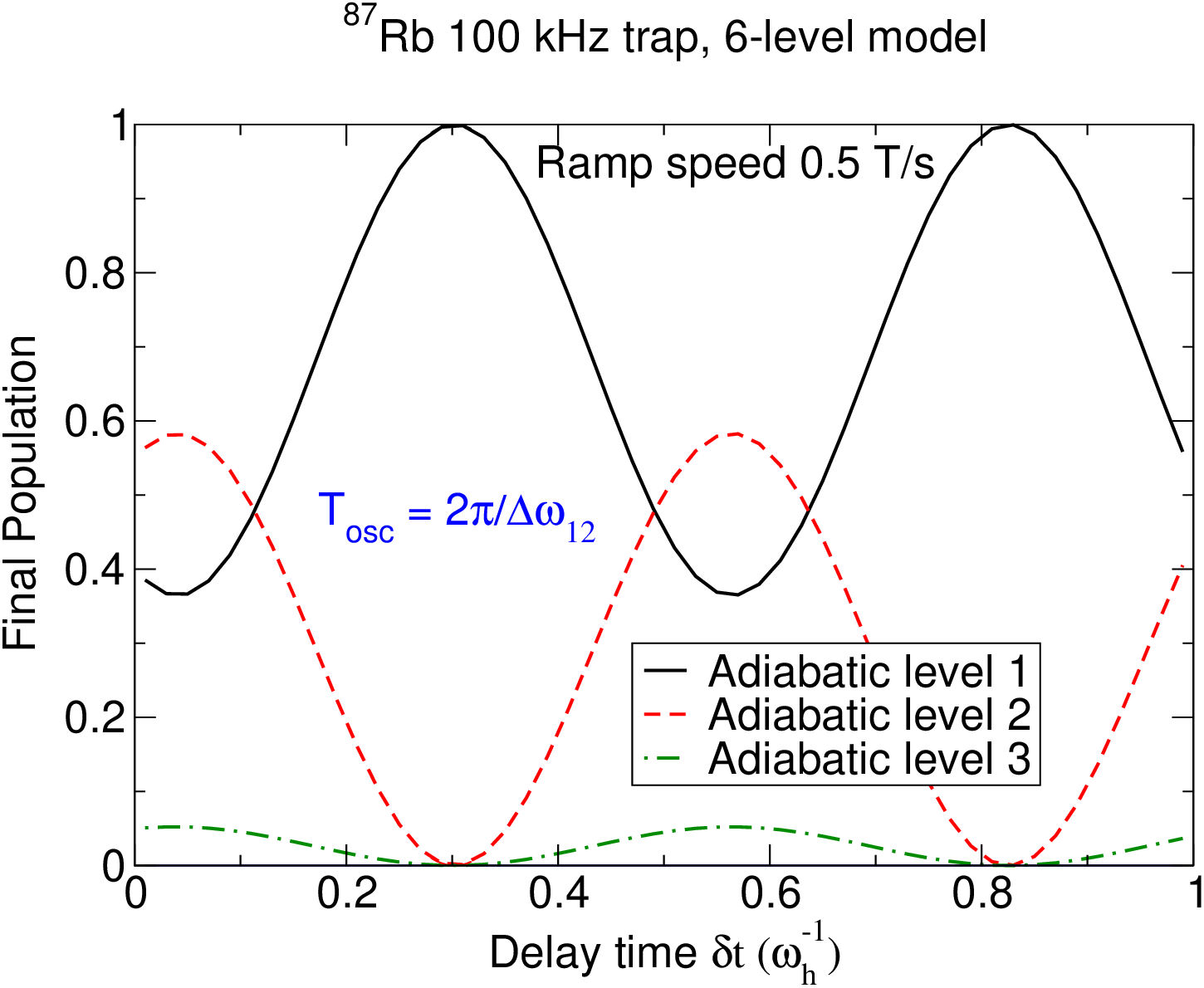}
\caption{Final state populations at long time $t\gg\omega_h^{-1}$ for the ramp sequence shown in Figure \ref{Fig3}.  Populations are shown for the adiabatic levels with the three lowest energies in Figure \ref{Fig3}.   The oscillation frequency is determined by the energy difference between the two lowest adiabatic levels (the approximate molecular resonance state and the trap ground state) during the hold time $\delta t$.}
\label{Fig4}
\end{figure}

A variety of different ramp types can be applied.  Figure~\ref{Fig3} gives an example of a double ramp, which first ramps the resonance level across threshold, then after a hold time $\delta t$, ramps it back in the opposite direction to be far above threshold at long time.  Figure~\ref{Fig4} shows how the final adiabatic populations at late time oscillate as $\delta t$ is changed.  The populations in the ground $v=0$ and first excited $v=1$ trap levels oscillate with a period $t=2\pi/\omega$ given by the difference $\hbar \omega$ between the energies of the (dressed) adiabatic trap ground level  and molecular resonance level during the hold interval.  For example, measuring the oscillation frequency for atom pairs in optical lattice cells for different hold magnetic fields could be used to measure $s_n=\partial{\epsilon_n}/{\partial B}$. 

The oscillations in Fig.~\ref{Fig4} for the case of two atoms in a lattice cell bear a similarity to observations on a $^{85}$Rb BEC using pulsed magnetic field ramps with a different time dependent configuration~\cite{Koehler03,Donley02}.  In these experiments the adiabatic bound state approached threshold from below and remained below threshold at all times.  However, the time-dependent ramps induced nonadiabatic transitions that resulted in the formation of dressed molecules and oscillations in the final atomic condensate populations at a frequency determined by the binding energy of the dressed molecular resonance state.  Reference~\cite{Borca03} applied a model of two atoms interacting in a trap to the condensate experiments of Ref.~\cite{Donley02}.  They adjusted their model parameters to account for the qualitative features of the experiment.  The next section describes the alternative approach introduced by Ref.~\cite{Mies00} to incorporate the many-body physics of a condensate into a model of trapped atom pairs.  This model has no adjustable parameters.

\section{Feshbach ramps in a BEC}\label{section3}

Mies {\it et al.}~\cite{Mies00} devised an approximate procedure to take into account ramping a time-dependent resonance level across threshold in a BEC.  They used the theory of atom pairs in the trap developed in Section~\ref{section2} and showed what changes must be made to account for the many-body effects in the condensate.  There are two essential effects that must be treated.  First, assuming that the background scattering length is not close to zero, the many-body mean field interactions cause the condensate size to expand to occupy a much larger volume than a single atom in the harmonic trapping potential, which would be confined to a volume of radius equal to $\sqrt{3}$ times the harmonic length for a single atom, $L_a=L_h/\sqrt{2}$.  In such a case, the condensate is characterized by the Thomas-Fermi radius
\be
L_{TF}=L_a \Re \,,
\ee
where
\be
\Re = \left ( \frac{15NA_{bg}}{L_a}\right )^\frac{1}{5} \,.
\ee

Since the decay width in Eq.~(\ref{CIwidth}), which is needed in Eq.~(\ref{Vnv2}),  is proportional to $\epsilon^{1/2}$, we need a criterion for picking a value of collision energy $\epsilon$ appropriate for the ground state of a condensate.  Ref.\cite{Mies00} used the result that the mean kinetic energy per atom in the condensate is~\cite{Fetter98,footnote1}
\be
\langle T_a \rangle = \frac{5}{2\Re^2} \ln{\left(\frac{\Re}{1.2683}\right)} \hbar \bar{\omega}_h
\label{Ta}
\ee
where $\bar{\omega_h}$ is the mean harmonic frequency of the confining potential.  Since the mean field potential exactly cancels the harmonic potential in the Thomas-Fermi approximation to make a flat effective potential for distances less than $L_{TF}$, we use for the ground state of the relative motion of a pair of condensate atoms the ground eigenstate of a box that has a zero point energy $\hbar \omega_b$ equal to $\langle T_a \rangle$ in Eq. (\ref{Ta}) (the kinetic energy of relative motion is equal to that of a single atom of the pair).   This gives
\be
 L_b = L_{TF} \left ( \frac{2}{5}\right )^\frac{1}{2} \frac{\pi}{\sqrt{\ln{\left(\frac{\Re}{1.2683}\right)}}} \,.
\label{BECBoxL}
\ee
Therefore, the harmonic trapping potential for the atom pair is replaced by a box potential that has the size $L_b$.  

The second effect of having a condensate is that the ground state is macroscopically occupied by $N$ atoms.  Each atom has $N-1$ atoms with which it can interact.  Since $N\gg1$, Ref.~\cite{Mies00}  introduced the factor $\sqrt{N}$ into the $V_{n,0}$ matrix element in Eq. (\ref{Vnvb}).  The resulting matrix element is proportional to $\sqrt{N/L_b^3}$.  Consequently, the Landau-Zener parameter in Eq. (\ref{ALZ}) is proportional to an atomic density. 

If we assume a $T=0$ condensate, there is no initial population in the excited trap levels with $v>0$.  Consequently, the dynamics is dominated by coupling of the resonance to the ground $v=0$ trap level, which is enhanced by the $\sqrt{N}$ factor, and we can ignore the excited levels, unless they become macroscopically populated in the course of the dynamics.  As discussed in Section~\ref{section3}, there are no transitions from $v=0$ to $v>0$ levels by a downward ramp, since these transitions are in a counterintuitive direction and are rigorously forbidden in a Landau-Zener model.  Thus, we might expect that the excited levels can be neglected for a downward ramp.  However, they can be populated when there is an upward ramp that moves population from the ground state to higher energy.  This is what happens in the MIT experiments~\cite{Inouye98,Stenger99} modeled in Ref.~\cite{Mies00}.  Even in this case the removal rate of atoms from the ground state depends only on $V_{n,0}$ and does not depend on the excited states.  We will show in the next Section that the predictions of this simple picture compare well with the results of proper many-body calculations of the system.

To calculate the time-dependent loss of condensate atoms when the Feshbach resonance is ramped across threshold from above or below, given that trap excitation can be neglected, all that is needed is to use the two coupled Eqs.~(\ref{CIeqs}) for the resonance state and the $v=0$ trap state, with the modified $V_{n,0}$  matrix element of Eq.~(\ref{Vnvb}) multiplied by $\sqrt{N}$ and using the box length from Eq.~(\ref{BECBoxL}).  These coupled equations describe an atom pair in the condensate.  If $N$ is the total number of atoms in the system, the number of atom pairs is $N/2$.  Of these atom pairs, after a downward ramp the fraction $p_0$ represents BEC atom pairs and the fraction $p_n=1-p_0$ represents molecules.  Consequently, the number of remaining condensate atoms is $Np_0$ and the number of molecules is $Np_n/2$, with $Np_n$ atoms combined into molecules.  Ref.~\cite{Mies00} showed that the Landau-Zener formula gives an excellent  account of the transition probabilities.   Let us call this method the LZ method. 

 It is straightforward to make a relatively minor correction to the theory to account for the nonlinear nature of the dynamics.   In the case of a BEC, we found that it is necessary to multiply the matrix element $V_{n0}^b$ in Eq.~(\ref{Vnvb}) by $\sqrt{N}$.  However, $N$ decreases dynamically during the course of the ramp as atoms are transferred out of the condensate.  This is readily treated by introducing the time dependent number of condensate atoms, $N|C_0(t)|^2$in Eqs.~(\ref{CIeqs}) for the condensate case, including only the $v=0$ ground state of the trap:
\begin{eqnarray}
i\hbar \dot{A_{n}}=&[\epsilon_{n}(B_{o})+\frac{\partial \epsilon 
_{n}}{\partial B}\frac{\partial B}{\partial 
t}t] A_{n}+V_{n,0} \sqrt{N |C_0|^2} C_{0} \nonumber \\
i\hbar \dot{C_{0}}=& \epsilon _{0} C_{0}+V_{n,0}\sqrt{N |C_0|^2} A_{n} \,.
\label{CIeqsMod}
\end{eqnarray}
This makes the dynamics nonlinear, as in the Gross-Pitaevsky equation, and corrects the LZ model in the right direction.  Let us call this the CI-BEC method.  Given the prescription in Eq.~(\ref{BECBoxL}), based on a physically motivated choice of collision energy, these equations have no adjustable parameters.

If the ramp is upward, the resonance energy ramps from below threshold to higher energies.  Even if the coupling is very strong between the ground trap (i. e., condensate) level and the resonance level, the coupling will be much weaker to the individual $v \ge 1$ levels because they are initially unoccupied.  So the crossings with the higher trap levels will have very small coupling and tend to be {\it diabatic}.  Thus there is little loss to the first few excited trap levels, and the population is carried to very high levels before it finally decays into the quasicontinuum of trap levels to produce hot atoms.  For the special case when only molecules are initially present without condensate atoms, a very simple expression~\cite{Mukaiyama03,Goral03} for the energy distribution of hot atoms can be worked out by simply assuming Fermi-golden rule decay of the resonance level $n$ at a time-dependent rate $\Gamma_n(\epsilon_n(t))/\hbar$ given by Eq.~(\ref{CIwidth}).

\section{Specific BEC calculations}\label{section4}

This Section will summarize the results of some specific calculations involving Feshbach ramps for $^{23}$Na, $^{87}$Rb, and $^{133}$Cs condensates.

\subsection{$^{23}$Na $F=1,M=+1$}\label{section4.1}

\begin{figure}[!htb]
\centering
\includegraphics[scale=0.3]{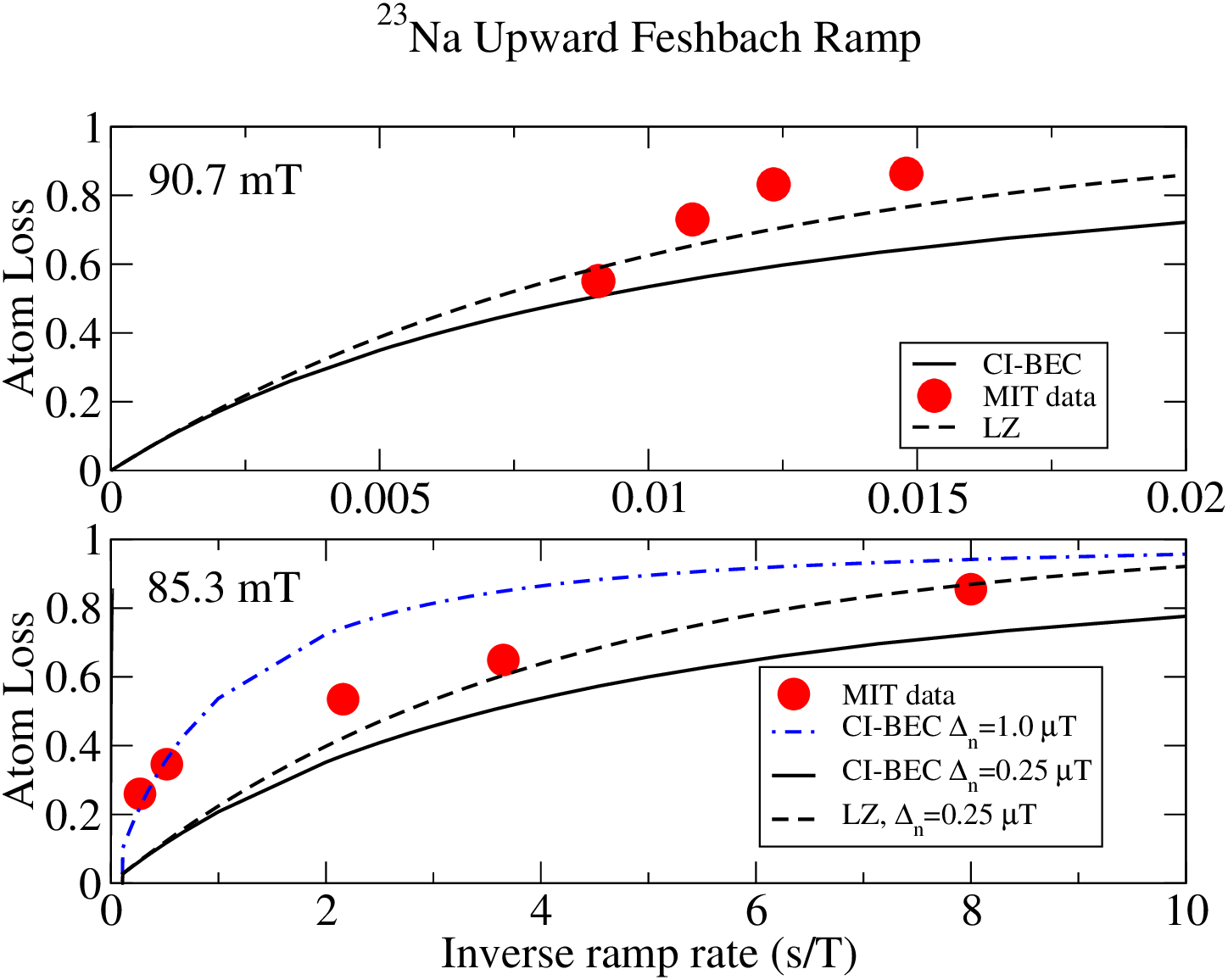}
\caption{Atom loss versus inverse ramp speed $|\partial{B}/\partial{t}|^{-1}$ for upward ramps of the 85.2 mT and 90.7 mT resonances in $^{23}$Na $F=1,M=+1$ collisions.  The figure compares the calculated LZ (dashed line) and CI-BEC (solid line) results with the experimental data points~\cite{Stenger99}.   In the case of the narrower 85.2 mT resonance, the CI-BEC results are shown for two different values of the resonance width, $\Delta_n=1$ $\mu$T and 0.25 $\mu$T.}
\label{Fig5}
\end{figure}

Figure~\ref{Fig5} shows LZ and CI-BEC calculations for the condensate atom loss $1-p_0$, using the 85.2 mT and 90.7 mT resonances for collisions of $^{23}$Na $F=1,M=+1$ atoms.  Mies {\it et al.}~\cite{Mies00} had previously applied the LZ method to these resonances to explain qualitatively the upward ramp experiments by the MIT group, which had measured the loss of condensate atoms associated with different ramp speeds $\partial{B}/\partial{t}$~\cite{Stenger99}.  The three model parameters  are $A_{bg}=3.32$ nm, $\Delta_n=0.1$ mT, and $s_n/h=52.9$ MHz/mT for the broad 90.7 mT resonance and $A_{bg}=3.38$ nm, $\Delta_n=1$ $\mu$T or 0.25 $\mu$T, with the same $s_n$ for the narrow 85.3 mT resonance.  The calculated width and estimated experimental width agree for the former, but differ by a factor of 4 for the latter~\cite{Stenger99}.  The calculations assumed a trap with the experimental mean trap frequency of $\bar{\omega_h}=2\pi(700$ Hz$)$ and $N=900000$ atoms.  The actual experimental optical trap was not spherical, but cigar-shaped.

Figure~\ref{Fig5} shows that the model gives the right order of magnitude of the atom losses.  If the renormalization of the matrix element for many-body effects described in Section~\ref{section3} were not taken into account, and a "bare" harmonic trap were used, then the  ramp speeds would differ by about three orders of magnitude to give the same loss fractions.  The CI-BEC calculations agree with the LZ calculations for fast ramp speeds (small inverse ramp speeds) where the atom losses are small, but give lower atom losses than the LZ calculations at slow ramp speeds, where the atom losses are large.  One would expect the LZ model to be better for fast ramps with small atom losses, since the initial condensate is less disturbed by the process.  The CI-BEC calculations for the fastest ramp speeds are closer to the experimental losses for the narrow 85.3 mT resonance when the theoretical width 1 $\mu$T is used instead of the estimated experimental width 0.25 $\mu$T of Ref.~\cite{Stenger99}.

Thus, it seems that the LZ formula in Eq.~\ref{ALZ} is useful for making order-of-magnitude estimates of atom losses using only three experimentally accessible parameters, $A_{bg}$, $\Delta_n$ and $\partial{B}/\partial{t}$, and two trap parameters, $\bar{\omega_h}$ and $N$, using the prescription in Eq.~(\ref{BECBoxL}) for the effective box size.  It is clear that proper many-body calculations are needed for quantitative studies.

\subsection{$^{87}$Rb $F=1,M=+1$}\label{section4.2}

Goral {\it et al.}~\cite{Goral03} carry out detailed calculations for the 100.76 mT resonance in $^{87}$Rb $F=1,M=+1$ collisions, comparing many-body calculations based on the method of K{\"o}hler {\it et al.}~\cite{Koehler02,Koehler03} with the LZ and CI-BEC models described here.  This is the same resonance discussed in Figs.~\ref{Fig1}-\ref{Fig4} for two atoms in an optical lattice cell.  Our many-body methods permit a clear way of counting the populations of the condensed and non-condensed atoms and the atoms combined in a molecular bound state.  The latter refers to the adiabatic, or "dressed", bound state of the two-body Hamiltonian, not the diabatic, or "bare", bound state that is only an intermediate theoretical construct and not an exact two-body eigenstate.   Thus, the method accurately counts the number of molecules that could be measured in an experiment in which they are captured.  We will summarize some of the results of Ref.~\cite{Goral03} here.  The key finding is that the CI-BEC model  gives remarkably good results when compared to many-body calculations.  

\begin{figure}[!htb]
\centering
\includegraphics[scale=0.3]{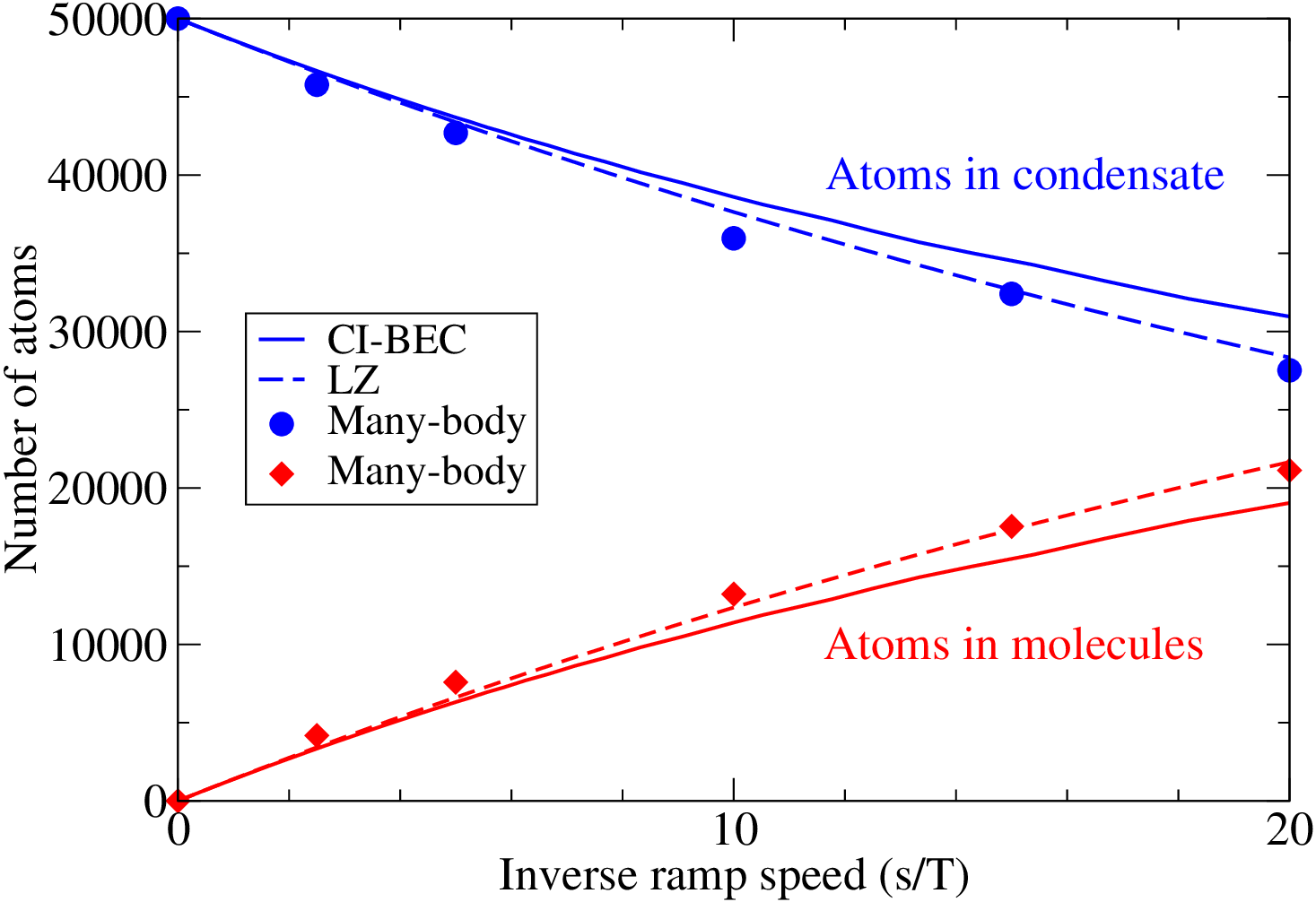}
\caption{Final populations versus inverse ramp speed $|\partial{B}/\partial{t}|^{-1}$ for a downward ramp of the 100.76 mT resonances in $^{87}$Rb $F=1,M=+1$ collisions in a 10 Hz trap with $N=50000$ atoms.  The figure compares the calculated LZ (dashed line) and CI-BEC (solid line) results with the many-body calculation for the remaining condensate atoms (solid circles) and atoms combined in the molecular resonance state (solid squares). }
\label{Fig6}
\end{figure}

\begin{figure}[!htb]
\centering
\includegraphics[scale=0.3]{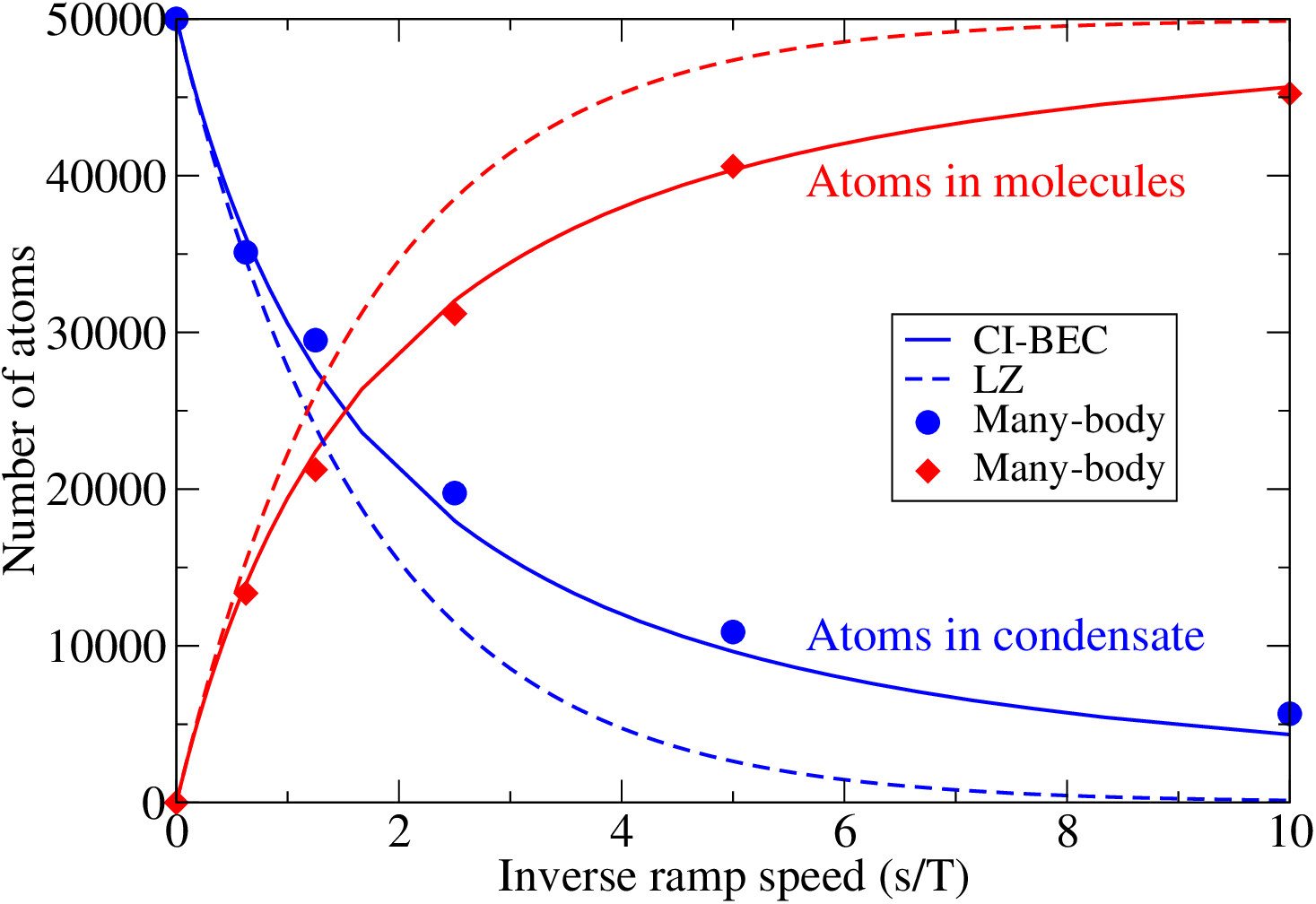}
\caption{Final populations versus inverse ramp speed $|\partial{B}/\partial{t}|^{-1}$ for a downward ramp of the 100.76 mT resonances in $^{87}$Rb $F=1,M=+1$ collisions in a 100 Hz trap with $N=50000$ atoms.  The figure compares the calculated LZ (dashed line) and CI-BEC (solid line) results with the many-body calculation for the remaining condensate atoms (solid circles) and atoms combined in the molecular resonance state (solid squares).}
\label{Fig7}
\end{figure}

Fig~\ref{Fig6} gives an example of calculations for a $\bar{\omega_h}=2\pi(10$ Hz$)$ trap with 50000 atoms.  The ramp starts with the resonance detuned far above threshold with all population in the condensate ground state and ends with the resonance tuned far below threshold.  The LZ and CI-BEC methods predict the number of remaining condensate atoms ($Np_0$) and number of atoms combined in molecules ($Np_n$) in good agreement with the many-body calculations~\cite{MBnote}.   In this case, the fractional depletion of atoms is modest, and the LZ and CI-BEC methods only differ slightly in their predictions.  Figure~\ref{Fig7} gives an example of a ramp of the same resonance in a $\bar{\omega_h}=2\pi(100$ Hz$)$ trap.   There is much more atom depletion and molecule formation in the 100 Hz trap than the 10 Hz trap for the same ramp speed, because the tighter trap gives larger matrix elements between the resonance state and the trap ground state.  The CI-BEC method agrees well with the many-body calculations, whereas the LZ method predicts too much atom loss from the condensate for the slower ramp speeds.  Thus, the very simple LZ model is good for estimating the magnitude of the ramp speed needed for conversion of atoms to molecules, although the actual efficiency of molecule production saturates less rapidly than the LZ model predicts.

\subsection{$^{133}$Cs $F=3,M=+3$}\label{section4.3}

Herbig {\it et al.}~\cite{Herbig03} have succeeded in using a Feshbach ramp to produce Cs$_2$ dimer molecules from a $^{133}$Cs condensate of $F=3,M=+3$ atoms trapped in an optical trap~\cite{Weber03}.  We have calculated the properties of the 2 mT resonance in these experiments using the theoretical model of threshold scattering developed by Leo {\it et al.}~\cite{Leo00} calibrated from the Feshbach spectroscopy experiments of Vuletic {\it et al.}~\cite{Vuletic00}.  We have also calculated the molecule production rates using the LZ, CI-BEC, and many body methods, using our calculated resonance parameters, $A_{bg}=8.63$ nm, $\Delta_n=0.5$ $\mu$T, and $s_n/h=7.98$ MHz/mT.  The $s_n$ value is in excellent agreement with the measured magnetic moment of the molecules~\cite{Herbig03}.  We will explain our calculations in full details elsewhere.   

The diabatic resonance level $|n\rangle$ has $(f m_f, \l m_{\l})=(44,42)$ molecular symmetry, where $f$ and $\l$ respectively represent spin and relative angular momentum with projections $m_f$ and $m_{\l}$.  This $g$-wave symmetry resonance state $|n\rangle$ is only very weakly coupled to the entrance channel $s$-wave collision. Our calculated resonance width is very sensitive to the short-range second-order spin orbit corrections to the effective spin-spin interaction, and is subject to error, possibly as much as a factor of two.  We find fairly good agreement in the limit of fast ramp speed with preliminary data from the Innsbruck group~\cite{Grimm03} on the number of molecules produced versus inverse ramp speed for a $\bar{\omega_h}=2\pi (11$ Hz$)$ trap with 50000 atoms.   

\begin{figure}[!htb]
\centering
\includegraphics[scale=0.3]{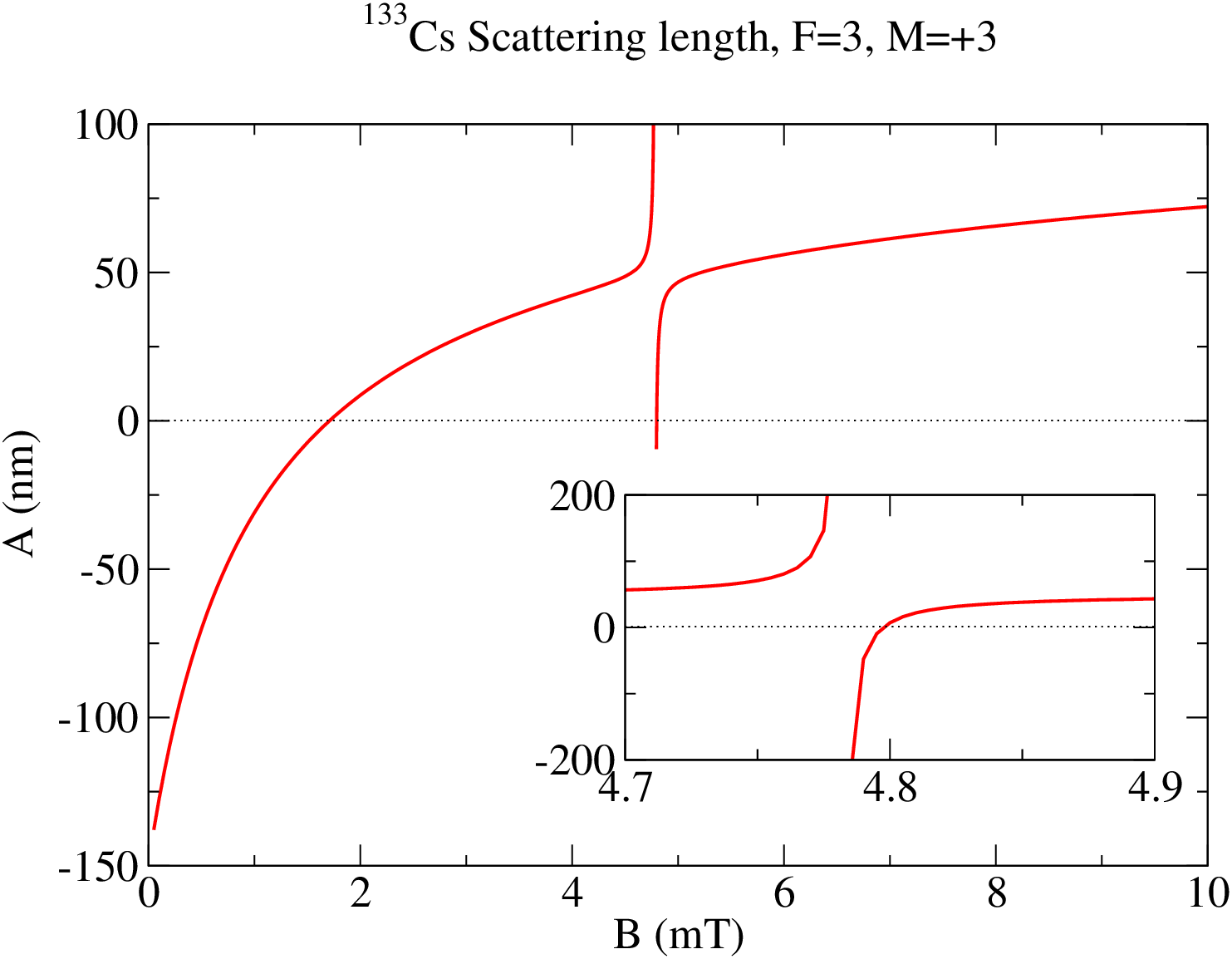}
\caption{Scattering length versus B, calculated using $\ell=0$ and 2 basis functions, with the inset showing the (44,22) resonance at 4.8 mT.  The (44,42) resonance at 2 mT studied by Herbig {\it et al.}~\cite{Herbig03} is missing, since $\ell=4$ basis functions were not included in this calculation.}
\label{Fig8}
\end{figure}

There is also a $^{133}$Cs resonance for $F=3,M=+3$ collisions near 4.8 mT with $(f m_f, \l m_{\l})=(44,22)$ molecular symmetry. Figure~\ref{Fig8} shows our calculated scattering length between 0 and 10 mT for collisions of atoms in this state, using the model of Leo {\it et al.}~\cite{Leo00} including $s$- and $d$-basis functions in the wavefunction expansion.   Atom losses due to ramping across this resonance were briefly described in the original condensate experiment~\cite{Weber03}, but with insufficient information for us to model.  It should also be possible to make molecules with this resonance.  The resonance parameters are $A_{bg}=47.9$ nm, $\Delta_n=15$ $\mu$T, $s_n/h=20.9$ MHz/mT.

\begin{figure}[!htb]
\centering
\includegraphics[scale=0.3]{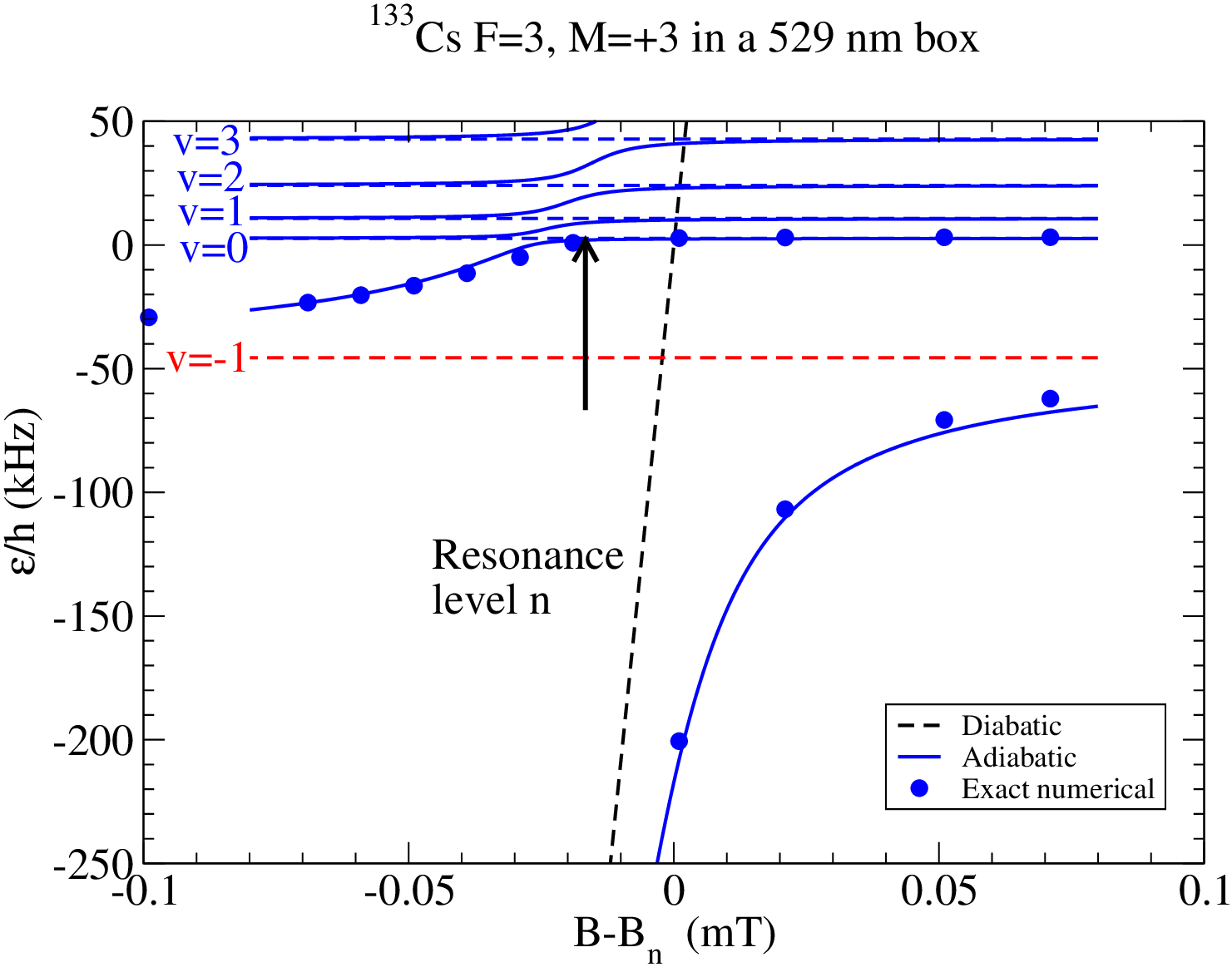}
\caption{Diabatic and adiabatic energy level versus $B-B_n$ for a CI model of a pair of $^{133}$Cs $F=3,M=+3$ atoms in a 529 nm box.  The points show the exact numerical eigenvalues using the full spin Hamiltonian in the model of Leo {\it et al.}~\cite{Leo00}.  The arrow shows the predicted location of the free-space threshold crossing, using the shift formula Eq.~(\ref{Shift}). }
\label{Fig9}
\end{figure}

The threshold crossing of the 4.8 mT resonance is significantly influenced by coupling to the last bound state of the background potential, which has a energy of only $\epsilon_{-1}/h=-45$ kHz.  Figure~\ref{Fig9} shows the bound states of relative motion of two $F=3,M=+3$ $^{133}$Cs atoms in a 529 nm box (529 nm = $10^4$ atomic units).  The lines show the diabatic and adiabatic energy levels calculated using the CI model described in Section~\ref{section2}.  The figure shows these are in good agreement with the numerical eigenvaules calculated from the full spin-dependent Hamiltonian and molecular potentials of the Leo {\it et al.}~\cite{Leo00} model.  The coupling of the resonance level $|n\rangle$ to the last bound state $|v=-1\rangle$ of the backgound potential is also included, using the calculated matrix element $V_{n,-1}/h=183$ kHz.  The strong avoided crossing of the $|n\rangle$ and $|v=-1\rangle$ levels is evident in the figure.   This avoided crossing is responsible for the large shift in the actual threshold crossing position.  The shift for this trap case is close to the free-space shift of $-1.10 \Delta_n$ predicted by Eq.~(\ref{Shift}) for our model parameters, $A_{bg}=47.9$ nm and $\bar{A}=5.11$ nm. 

\begin{figure}[!htb]
\vspace{1cm}
\centering
\includegraphics[scale=0.3]{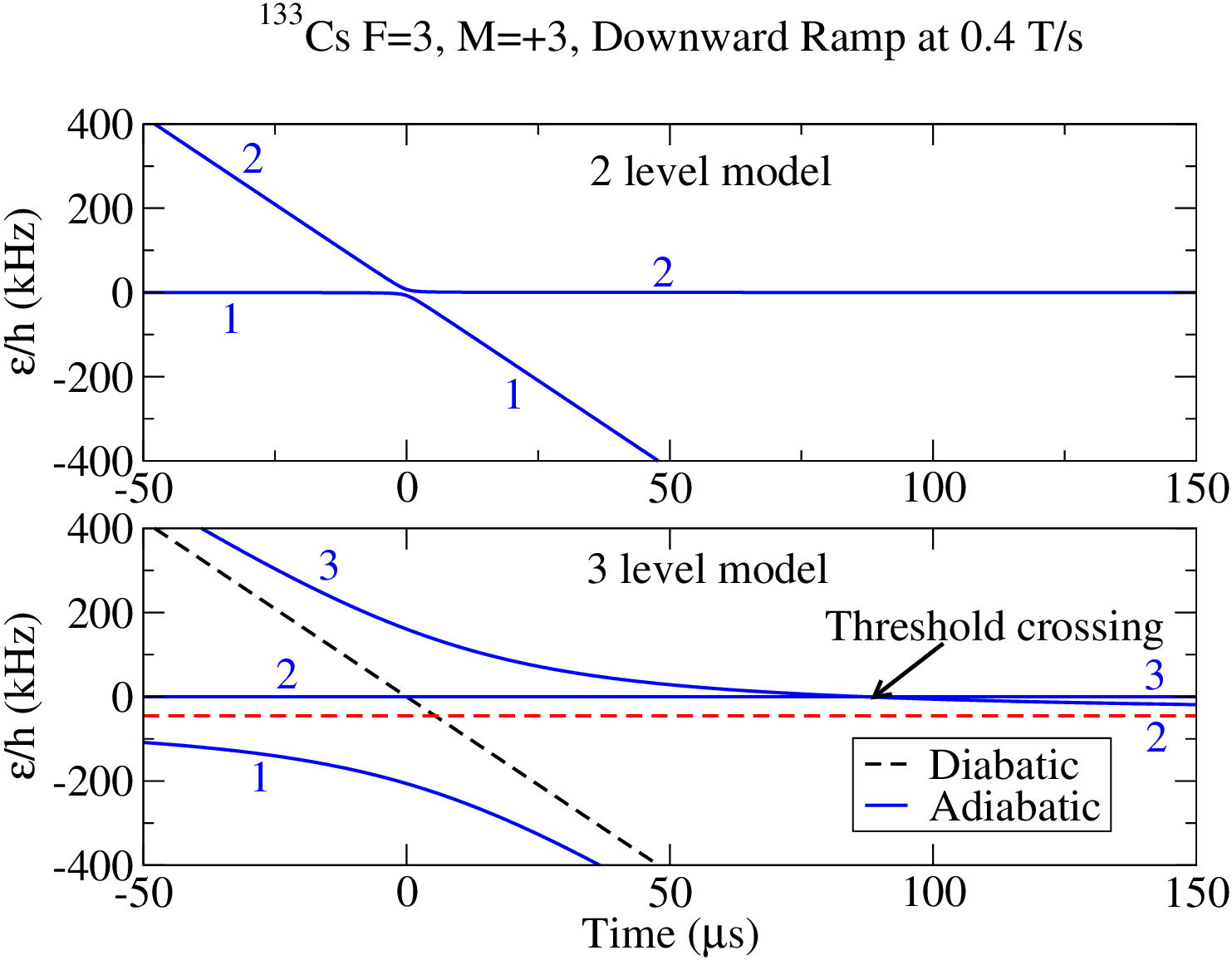}
\caption{Energy levels of the LZ Hamiltonian for a $^{133}$Cs $F=3,M=+3$ condensate in a $\bar{\omega_h}=2\pi(20$ Hz$)$ trap with 12500 atoms, for the cases of  two-level and three-level models for a downwards ramp near 4.8 mT with speed 0.4 T/s.  The dashed lines show the diabatic levels.  The numbers label the adiabatic levels in order of increasing energy.  The three-level model also includes the last bound level  $|v=-1\rangle$ of the open channel potential at -45 kHz.  The strong interaction between the resonance level $|n\rangle$ and $|v=-1\rangle$ leads to a strong avoided crossing that changes the nature of the threshold crossing, shifting it to 88 $\mu$s later in time for the ramp speed used here. In the three-level model the adiabatic level 2 evolves from its initial character as the trap ground state $|v=0\rangle$ to its final character as the $|v=-1\rangle$ molecular state. }
\label{Fig10}
\end{figure}

Figure~\ref{Fig10} shows the adiabatic energy levels for the LZ Hamiltonian matrix using a two-level model with $|n\rangle$, $|v=0\rangle$ in the basis and a three-level model with $|v=-1\rangle$ also in the basis.  In the two-level model there is an avoided crossing centered on the time that the resonance level $|n\rangle$ crosses threshold, and adiabatic evolution takes the initially populated trap ground state atoms in level $|v=0\rangle$ to molecular resonance level $|n\rangle$ population for times long after the crossing.  By contrast, the strong interaction with the last bound state in the three-level model shifts the threshold crossing to later time, and adiabatic evolution would lead to producing molecules in the $|v=-1\rangle$ molecular state instead of in the diabatic resonance state $|n\rangle$.

\begin{figure}[!htb]
\centering
\includegraphics[scale=0.3]{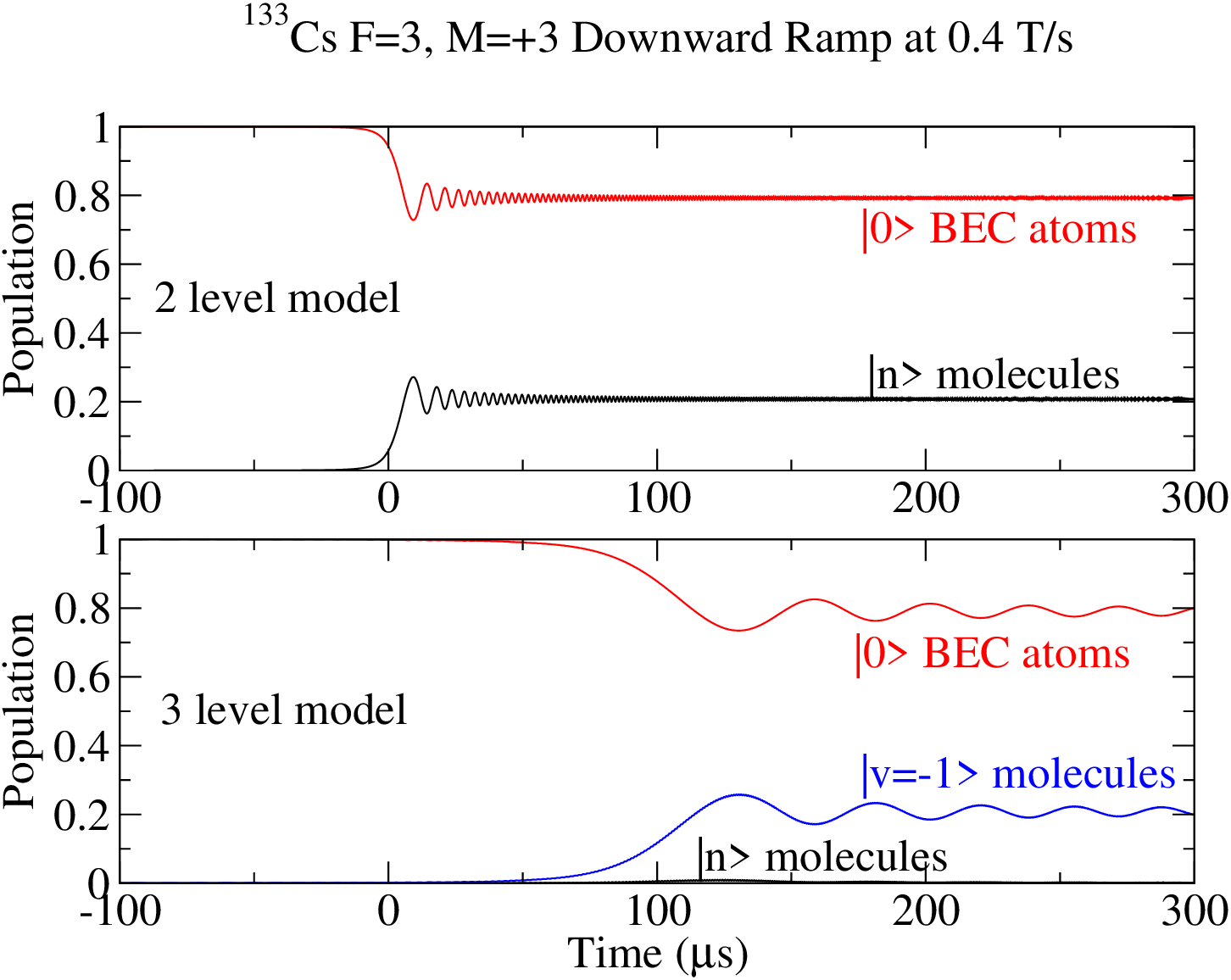}
\caption{Calculated diabatic populations using the CI-BEC equations for the two- and three-level models shown in Fig.~\ref{Fig10}. The inclusion of the last bound state profoundly changes the detailed dynamical evolution, although the final state populations of condensate atoms and molecules is the same in both cases and is close to the value given by the LZ formula.  In the three-level case, the population of the resonance state $|n\rangle$ always remains small, and is barely visible in the figure.}
\label{Fig11}
\end{figure}

Figure~\ref{Fig11} shows the time-dependent populations calculated using the CI-BEC model for the case shown in Fig.~\ref{Fig10}.  Since the LZ parameter describing the loss only depends on the $V_{n0}$ matrix element, the LZ model predicts the same final loss of initial atomic population for both the two- and three-level models.  However, the detailed population dynamics, and the nature of the final molecules produced, are profoundly different between the two models.  When we consider the effect of the last bound state, the actual transfer of population from atoms to molecules occurs near the time of actual crossing, shifted by the strong interaction with the last bound state, and results in the formation of molecules in the last bound state $|v=-1\rangle$ of the background potential, not formation of resonance state molecules $|n\rangle$.  It is interesting to note that in the three-level model, even at intermediate times near the resonance crossing, there is never significant population in the resonance state $|n\rangle$.

In closing, we note that experiments could also be done for $^{133}$Cs $F=3,M=+3$ atom pairs trapped in optical lattice cells, as described in Section~\ref{section2}.   Similar qualitative behavior to that described in Figs.~\ref{Fig10} and \ref{Fig11} should occur.   The dynamics would contrast sharply from that described in relation to Figs.~\ref{Fig1} and \ref{Fig2} for $^{87}$Rb $F=1,M=+1$, for which  $A_{bg}$ is much closer to $\bar{A}$, and the last bound state is much more deeply bound by $\epsilon_{-1}/h=-23$ MHz~\cite{Goral03}.  Interaction of $|n\rangle$ with $|v=-1\rangle$ results in only a slight modification of the threshold crossing, and would not significantly change the results in Fig.~\ref{Fig2} for  ramps that stop before allowing $\epsilon_n$ to cross $\epsilon_{-1}$.  Of course, if the ramp should carry $\epsilon_n$ to be close to or to cross $\epsilon_{-1}$, the mixing of these levels leads to the very interesting physics that has been exploited by D\"urr {\it et al.}~\cite{Durr03} in making molecules from $^{87}$Rb condensate atoms.

\section{Summary}

We have presented a review and two extensions of the Feshbach ramp model of Mies {\it et al.}~\cite{Mies00} for describing the conversion of pairs to trapped atoms to molecules by ramping the energy of a Feshbach resonance state from above to below a collision threshold.  The review outlines the key ideas behind the method.  The extensions are given by in Eqs.~(\ref{Vnvh})-(\ref{Vnvb}) and ~(\ref{CIeqsMod}).  The former equations express the matrix elements using the Wigner-law form of the coupling, and the latter Eq.~(\ref{CIeqsMod}) introduces nonlinear dynamics for the case of condensate evolution.  Given the number of atoms in the trap and the mean trap frequency, the model is fully characterized in terms of only three nonadjustable resonance scattering parameters which can be measured or calculated: the background scattering length, the width and the magnetic moment of the resonance state.  The model can be supplemented by adding the coupling matrix element to the last bound state in the background potential if necessary.  Given the ramp speed, the model predicts the loss of atoms from the initially populated ground state of the trap and the conversion efficiency to molecules.  Simple estimates based on the Landau-Zener curve crossing formula give the correct order of magnitude of atom loss when compared to experimental cases.  The CI-BEC equations compare quite favorably with the predications of many-body theory for condensates.

The model can be extended to other cases.  It could be readily adapted to one- or two-color photoassociation.  This can be treated by a Feshbach resonance formalism that includes the decay of the excited state by spontaneous emission.  In the two-color case, it is possible to suppress this decay by using an interference mechanism~\cite{Julienne98}.  The model also applies to two Fermions with different spins, and could be applied without modification to pairs of fermionic atoms in single trapping cells.  It is possible that the model could be adapted to quantum degenerate mixed fermionic gases.  One important physical effect that the model leaves out is the role of other collisions during the ramp process.  For example, if a third atom approached the colliding pair of atoms, a resonance-modified three-body process occurs.  There clearly is a rich variety of phenomena that can be studied based on manipulating resonant scattering processes in optical lattices or quantum degenerate gases.  It is hoped that models like the one presented here will help stimulate thinking about these phenomena.

\acknowledgments
We thank the Office of Naval Research for partial support.  This research was also supported
by a University Research Fellowship of the Royal Society (T.K.).


\begin{thebibliography}{99}
\bibitem{MBnote} Since it was impractical to run the many-body calculations to the same large detunings as the CI-BEC calculations, small remaining time-dependent oscillations contribute to a small amount of "noise" in the many-body results.
\end{thebibliography}
\end{document}